\documentclass[aip, apl, superscriptaddress, reprint]{revtex4-1}
\usepackage{amsmath}
\usepackage{graphicx}
\usepackage{amssymb}
\usepackage{url}
\usepackage{bm}

\begin{document}


\title{High-Q silicon photonic crystal cavity for enhanced optical nonlinearities} 



\author{Ulagalandha Perumal Dharanipathy}
\affiliation{Institut de Physique de la Mati\`{e}re Condens\'{e}e, Ecole Polytechnique F\'{e}d\'{e}rale de Lausanne (EPFL), CH-1015 Lausanne, Switzerland}

\author{Momchil Minkov}
\email[]{momchil.minkov@epfl.ch}
\affiliation{Laboratory of Theoretical Physics of Nanosystems, Ecole Polytechnique F\'{e}d\'{e}rale de Lausanne (EPFL), CH-1015 Lausanne, Switzerland}

\author{Mario Tonin}
\affiliation{Institut de Physique de la Mati\`{e}re Condens\'{e}e, Ecole Polytechnique F\'{e}d\'{e}rale de Lausanne (EPFL), CH-1015 Lausanne, Switzerland}

\author{Vincenzo Savona}
\affiliation{Laboratory of Theoretical Physics of Nanosystems, Ecole Polytechnique F\'{e}d\'{e}rale de Lausanne (EPFL), CH-1015 Lausanne, Switzerland}

\author{Romuald Houdr\'{e}}
\affiliation{Institut de Physique de la Mati\`{e}re Condens\'{e}e, Ecole Polytechnique F\'{e}d\'{e}rale de Lausanne (EPFL), CH-1015 Lausanne, Switzerland}

\date{\today}

\begin{abstract}
We fabricate and experimentally characterize a $H0$ photonic crystal slab nanocavity with a design optimized for maximal quality factor, $Q = 1.7$ million. The cavity, fabricated from a silicon slab, has a resonant mode at $\lambda = 1.59 \mathrm{\mu m}$ and a measured $Q$-factor of $400,000$. It displays nonlinear effects, including high-contrast optical bistability, at a threshold power among the lowest ever reported for a silicon device. With a theoretical modal volume as small as $V = 0.34(\lambda/n)^3$, this cavity ranks among those with the highest $Q/V$ ratios ever demonstrated, while having a small footprint suited for integration in photonic circuits.
\end{abstract}


\maketitle 

Photonic crystal (PhC) nanocavities are promising building blocks of future integrated photonic circuits\cite{Nozaki2010, Nozaki2012, Takahashi2013, Notomi2010}. Considerable effort has been devoted in the last decade to the optimization of these structures. In particular, several studies have been aiming at developing designs with the highest quality factor $Q$, combined with the smallest modal volume $V$ as the nonlinear optical response of these devices -- as well as the Purcell effect and radiation-matter coupling -- are enhanced as $Q$ increases and $V$ decreases. \cite{Andreani1999,Barclay2005, Englund2005a,Yoshie2004} Few specific designs, where the nanocavity originates as a local defect in a PhC waveguide, have reached measured $Q$-factors exceeding one million \cite{Song2005, Tanabe2007, Taguchi2011, Sekoguchi2014}. More recently, \cite{Minkov2014} by combining a fast simulation tool to a genetic optimization algorithm, we have systematically optimized the three most widespread cavity designs -- the $H0$, $H1$, and $L3$ designs -- to theoretical quality factors largely exceeding one million, using only shifts in the positions of a few neighboring holes. Compared to previous optimization attempts,\cite{Akahane2005, Akahane2003, Nomura2010, Notomi2010, Takagi2012, Tanabe2007a, Kuramochi2013, Zhang2004} the $Q$-factors of these cavities were thus improved sometimes by more than one order of magnitude, while their modal volumes were not significantly increased when compared to the unoptimized designs.

\begin{figure*}
\includegraphics[width = 0.9\textwidth]{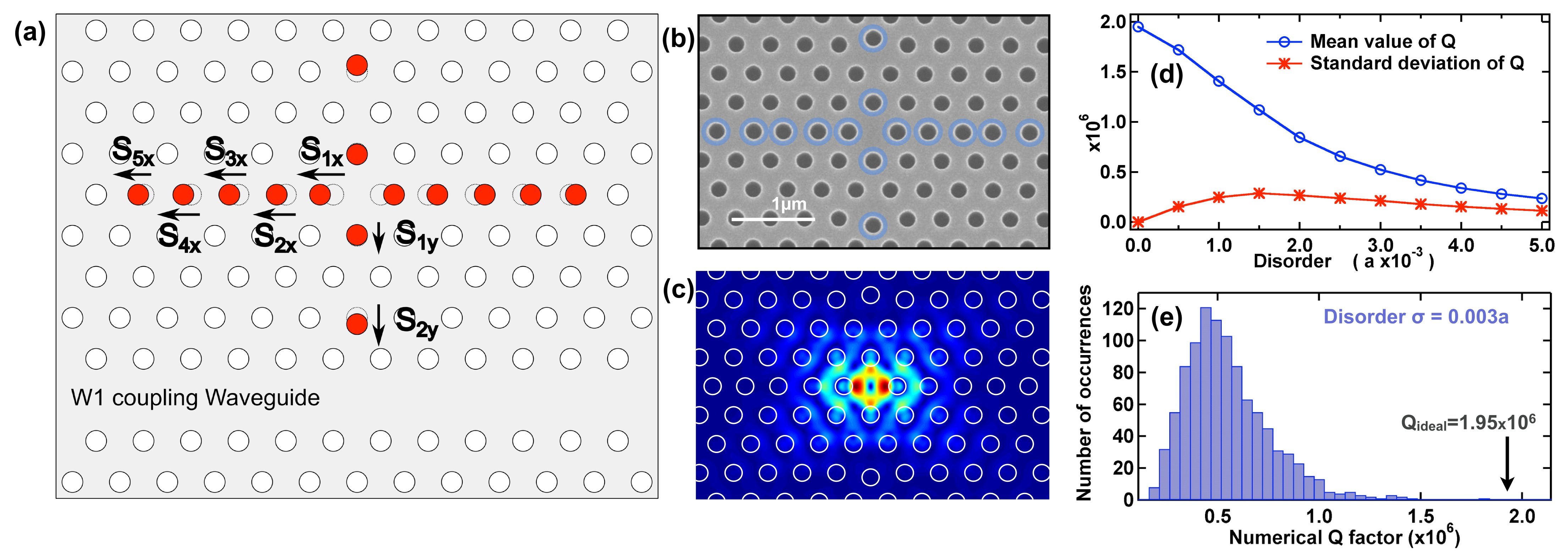}%
\caption{\label{fig1} (a): Schematic of the proposed design, highlighting (in red) the holes whose positions have been optimized, with the displacement parameters correspondingly labelled. The coupling W1 waveguide, at a distance from the cavity, is shown. (b): Scanning electron micrograph of one of the fabricated cavities, with the displaced holes encircled in blue. (c): Electric field distribution in the cavity as computed with the guided-mode expansion. (d) Mean and standard deviation of $Q$ in the presence of random structural disorder, as a function of the disorder magnitude $\sigma$, computed from 1000 simulated disorder realizations for each $\sigma$. (e): Histogram of the computed quality factor for $\sigma = 0.003a$; the ideal $Q$-factor without disorder is indicated.}
\end{figure*}

Here, we fabricate and characterize the optimal $H0$ (also known as ``point-shift'' or ``zero-cell'') design derived in Ref. \onlinecite{Minkov2014}. The original $H0$ design  \cite{Zhang2004,Nozaki2006,Nomura2010} has a mode volume $V=0.23(\lambda/n)^3$ -- i.e. the smallest mode volume among PhC slab nanocavities.\footnote{Here, we use the definition $V= \frac{\int\varepsilon(\bm{r})|\bm{E}(\bm{r})|^2d^3r}{\mbox{max}\left[\varepsilon(\bm{r})|\bm{E}(\bm{r})|^2\right]}$}$^,$\cite{Vuckovic2001} Its quality factor had previously been optimized to a moderate theoretical $Q = 280,000$, while in our recent work \cite{Minkov2014} we reached a theoretical value close to two million using similar variational parameters. The optimal cavity design that we consider is illustrated in Fig. \ref{fig1}(a). The thickness of the PhC slab is $220\mathrm{nm}$ while the radius of each hole is $0.25a$ ($a$ is the lattice constant), and the refractive index is $n = 3.46$. The basic $H0$ design \cite{Zhang2004,Nozaki2006} consists of two holes shifted away from their original positions by an amount $S_{1x}$ (see Fig. \ref{fig1}(a)). The optimized design was obtained by allowing for four more shifts of neighbouring holes along the x-axis and two shifts along the y-axis, as shown in Fig. \ref{fig1}(a). The objective function of optimization was the cavity quality factor $Q$, while reasonable restrictions were imposed on the magnitudes of the shifts in order to limit the variations in modal volume and to exclude hole overlap. The main computational tool employed in this work for the simulation of a nanocavity structure is the guided-mode expansion (GME),\cite{Andreani2006} that was used both for the optimization and for the disorder analysis presented below. The optimal $H0$ parameters are: $S_{1x}=0.280a$,  $S_{2x}=0.193a$, $S_{3x}=0.194a$, $S_{4x}=0.162a$, $S_{5x}=0.113a$, $S_{1y}=-0.016a$, (i.e. shifted inward) and $S_{2y}=0.134a$, bringing a GME-computed quality factor $Q_{ideal} = 1.96 \times 10^6$. The computed mode profile (Fig. \ref{fig1}(c)) resembles that of the basic design and, most importantly, the modal volume remains extremely small: $V = 0.34(\lambda/n)^3$. We additionally simulated the optimal structure using a 3D finite-difference time-domain method,\cite{Oskooi2010} which confirms the GME-computed volume and gives $Q_{ideal} = 1.7 \times 10^6$ -- in good agreement with the GME value.

In order to assess the robustness of this design, we modelled random disorder in the form of fluctuations in the hole positions and radii,\cite{Hughes2005,Koenderink2005,Portalupi2011,Hagino2009,Minkov2013} drawn from a Gaussian random distribution with zero mean and standard deviation $\sigma$. In Fig. \ref{fig1}(d), we plot the mean value of the quality factor and its standard deviation in the presence of disorder as a function of $\sigma$, where each point was computed based on 1000 random disorder realizations. Disorder reduces the $Q$-factor on average, as expected,\cite{Portalupi2011, Hagino2009, Minkov2013} but nonetheless very high quality factors for a reasonable fabrication disorder magnitude are predicted, proving the robustness of this design in terms of practical applications. Fig. \ref{fig1}(e) shows a histogram of the probability distribution of the $Q$-values for  $\sigma = 0.003a$, which is a reasonable estimate of the largest fluctuations introduced in our fabrication process\cite{LeThomas2011}, and is consistent with the experimental results below. 

\begin{figure*}
\includegraphics[width = 0.8\textwidth]{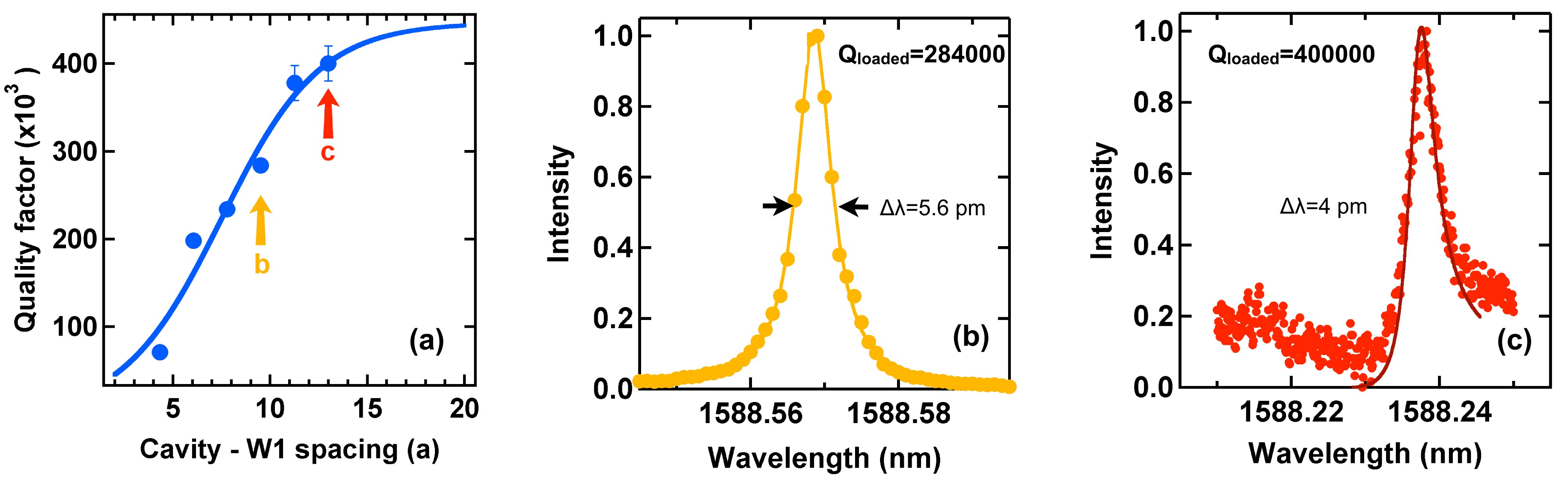}%
\caption{\label{fig2}(a): Change in the measured $Q$-factor as the distance $D$ from the side-coupling W1 waveguide is increased. The blue line shows the best fit of the experimental data to the model $Q^{-1} = Q_{UL}^{-1} + C\mathrm{exp}(-\alpha D)$, indicating an unloaded $Q$-factor $Q_{UL} = 450,000$. (b) and (c): Spectra (normalized to the maximum intensity) of the emission from the membrane surface measured for two values of the cavity-waveguide distance, indicated by arrows in panel (a). The experimental data are fitted with a Lorenzian curve in panel (b) and a Fano curve in panel (c), and the extracted full width at half maximum results respectively in loaded quality factors $Q_L = 284,000$ and $Q_L = 400,000$.}
\end{figure*}

\begin{figure}[h!]
\includegraphics[width = 0.45\textwidth, trim = 0in 0in 0in 0in, clip = true]{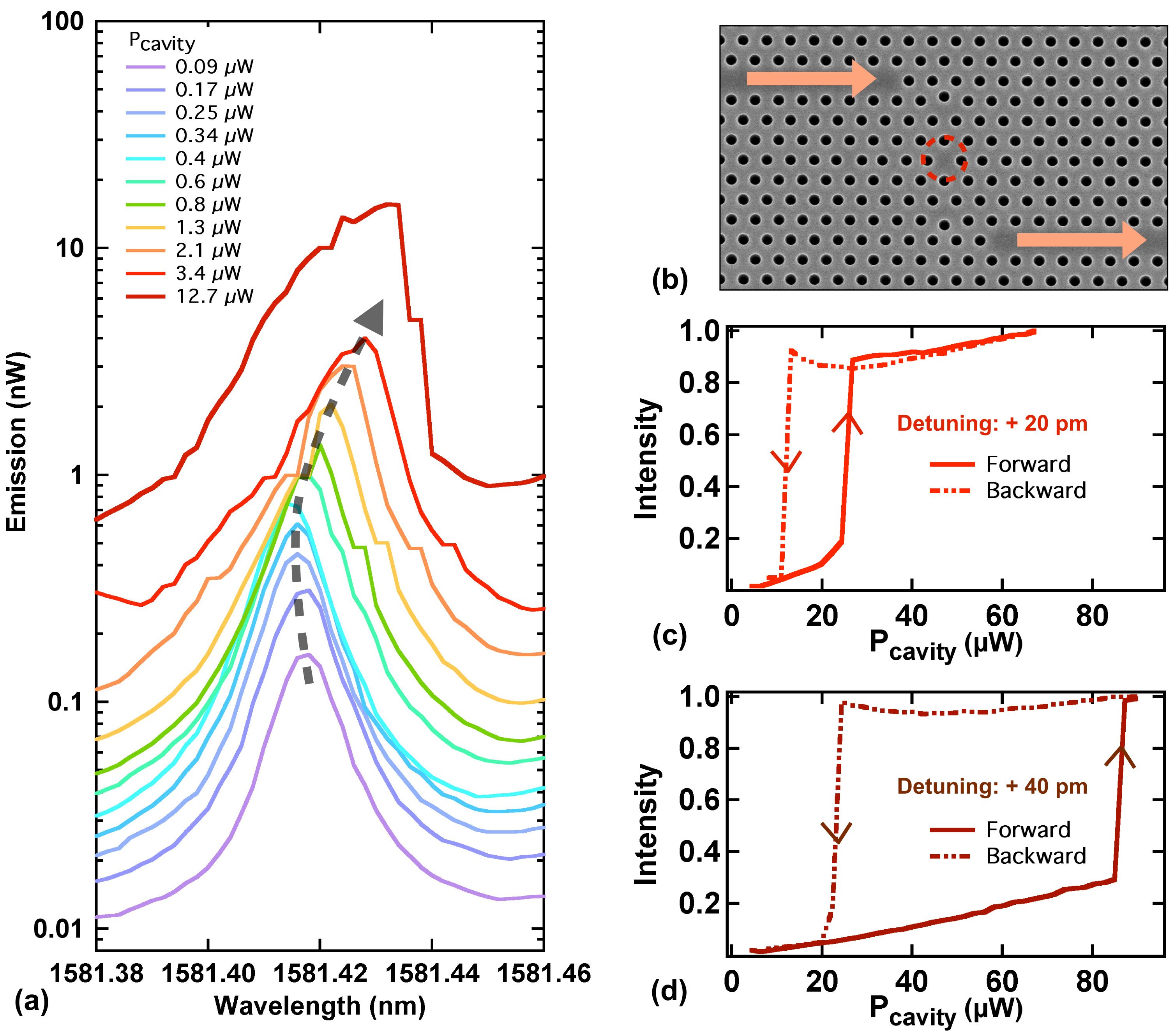}%
\caption{\label{fig3}(a): Measured emission spectrum from the cavity, as the input power is increased. The dashed line is a guide to the eye to indicate the change in the resonance wavelength. (b): Scanning electron micrograph of the cross-coupled cavity structure used to measure optical nonlinearities. (c) and (d): Hysteresis plots, respectively for an excitation wavelength red-shifted by $20\mathrm{pm}$ and $40\mathrm{pm}$ from the cavity resonance.}
\end{figure}

Several nanocavities were fabricated following the optimal design with $a = 435\mathrm{nm}$, on a Silicon-On-Insulator wafer, which consists of a $220\mathrm{nm}$ thick silicon layer and a $2\mathrm{\mu m}$ thick silica ($\mathrm{SiO_2}$) layer on a silicon substrate. The photonic crystal pattern is defined with electron beam lithography (VISTEC EBPG5000) on an electro-sensitive resist (ZEP520) and the developed pattern is further transferred into the silicon layer with an inductively coupled plasma (ICP) AMS200 dry etcher with a $\mathrm{SF_6}$ and $\mathrm{C_4F_8}$ gas mixture. The last step is the removal of the sacrificial $\mathrm{SiO_2}$ layer with buffered HF (BHF) wet etching. Coupling of continuous-wave monochromatic light into the cavity was performed in a standard end-fire set-up with lensed fibres, adiabatically tapered ridge waveguides and photonic crystal W1 waveguides. The cavity couples either in a side-coupling (Fig. \ref{fig1}(a)) or in a cross-coupling (Fig. \ref{fig3}(b)) scheme, and was characterized by a different distance to the waveguide $D = n\frac{\sqrt{3}}{2}a, \, n = 5, \dots 15$. When measuring the $Q$-factors, the input light power was lowered until optical nonlinearities (see below) vanished and the device operated in a regime of linear response. Cavity emission spectra are shown in Fig. \ref{fig2}(b) and (c) for $n = 11$ and $n = 15$, respectively. In panel (c), the coupling of light into the cavity is very weak, making the signal almost comparable to the noise, which is why we used a Fano fit instead of a Lorentzian (the former describes the spectral response of a resonance in a continuous background). Fig. \ref{fig2}(a) shows the change in measured (loaded) $Q$-factor, for the side-coupled cavities, as $D$ is increased. The variation is due to the coupling waveguide that acts as an additional loss channel for the cavity.\cite{Akahane2005} The error bars of the data points (only visible for the last two points on the scale of the plot) come from the uncertainty in the Lorentzian/Fano fits, and do not take into account the variation in $Q$ values that is expected among different cavities due to disorder. This, as suggested by Fig. 1(d) and (e), is expected to be much larger than the measurement error. The maximum measured value of $Q = 400,000$ was obtained for a coupling distance $n = 15$. The data in Fig. \ref{fig2}(a) suggest clearly that, at $n = 15$, the coupling waveguide still affects the measured $Q$-factor. A conservative way of extrapolating the unloaded $Q$-factor  consists in assuming an exponential decay with distance of the cavity-waveguide coupling. More precisely, we assume $Q^{-1} = Q_{UL}^{-1} + C\mathrm{exp}(-\alpha D)$. A fit of the measured $Q$-values (with $C$, $Q_{UL}$, and $\alpha$ as free parameters), as plotted in Fig. \ref{fig2}(a), yields $Q_{UL} = 450,000$, which should be taken as a lower bound to the actual unloaded $Q$-factor. This value is in very good agreement with the maximum in the histogram of Fig. \ref{fig1}(e), computed for a disorder amplitude $\sigma = 0.003a$, which is a very reasonable estimate of the largest fluctuations introduced in the fabrication process.\cite{LeThomas2011} Finally, we note that Fig. \ref{fig2}(a) also shows that at short distances, where the $Q$-factor is still very high ($\approx 100,000$), losses are fully dominated by coupling into the waveguide channel, which highlights the potential for photonic applications. 

The simulated modal volume $V=0.34(\lambda/n)^3$ is a reliable estimate of the corresponding quantity for the fabricated structure \cite{Vignolini2010,Burresi2010}. The present nanocavity thus ranks among those with the highest $Q/V$ ratio ever reported.\cite{Nozaki2010, Taguchi2011, Akahane2003, Takagi2012, Barclay2005, Deotare2009, Takahashi2008, Asano2006}
In the present work, we have additionally investigated an alternate optimal design \cite{Minkov2014} characterized by a smaller modal volume, that was obtained by introducing a stricter upper bound $S_{1x} < 0.25a$ in the optimization procedure. The design has an ideal GME-computed $Q$-factor $1.05 \times 10^6$ (FDTD: $1.0 \times 10^6$) and a smaller modal volume $V = 0.25(\lambda/n)^3$. The shifts defining this design are as follows: $S_{1x}=0.216a$,    $S_{2x}=0.103a$,    $S_{3x}=0.123a$,  $S_{4x}=0.004a$,  $S_{5x}=0.194a$, $S_{1y}=-0.017a$,   $S_{2y}=0.067a$. This cavity was also fabricated and experimentally characterized, and a maximum $Q$-factor of $260,000$ was measured. 

The $Q/V$ ratio is a measure of the enhancement of optical nonlinearities produced by a cavity.\cite{Uesugi2006, Barclay2005, Notomi2005} To examine the nonlinear spectral properties of our design, in Fig. \ref{fig3}(a) we show the emission spectrum, measured under continuous-wave resonant excitation, of a cavity with measured $Q$-factor $Q=150,000$ in the cross-coupling configuration, at varying intra-cavity power. This configuration allows for a rough estimate of the power coupled into the PhC region where the cavity is located \cite{Notomi2010, Zhang2013}. For a given input and output power, we define the transmission coefficient $T = P_{output}/P_{input}$. For a symmetric system (Fig. \ref{fig3}(b)), the power available in the cavity region is then $P_{cavity} = \sqrt{T}P_{input}$ -- i.e. it depends linearly on the actual input power -- which can also be rewritten, using $T$ from the previous expression, as $P_{cavity} = \sqrt{P_{input}P_{output}}$. In particular, for an input power from the laser $P_{input} = 4\mathrm{\mu W}$, we measure a transmitted power $P_{output} \approx 2\mathrm{nW}$ in the detector after the ridge waveguide at the cavity resonance wavelength. According to the above equation, for these values $P_{cavity} = 0.09\mathrm{\mu W}$, which sets the proportionality factor $\sqrt{T}$ between $P_{cavity}$ and $P_{input}$ in our measurement. As shown in panel (a) of the figure, at low power, a slight blue shift of the cavity mode is observed and attributed to free carrier dispersion.\cite{Barclay2005} Starting at $P_{cavity} = 0.6\mathrm{\mu W}$, heating due to nonlinear absorption, and optical-Kerr nonlinearity result in a redshift. At higher input powers, a drop in the spectral response on the red side of the resonance indicates the onset of optical bistability. To characterize this bistable behaviour, we sweep the input power and record the steady-state emission intensity. A clear hysteresis with a large contrast and very low power threshold is observed in Figs. \ref{fig3}(c) and (d), where the input laser is respectively detuned by $20\mathrm{pm}$ and $40\mathrm{pm}$ above the cavity resonance. Switching power ratios $P_{up}/P_{down}$ of respectively 2.0 ($P_{up} = 26\mathrm{\mu W}$ and $P_{down} = 13\mathrm{\mu W}$) and 4.5 ($P_{up} = 90\mathrm{\mu W}$ and $P_{down} = 20\mathrm{\mu W}$), and a contrast above 70$\%$ are obtained, demonstrating robust and controllable bistable behaviour. The present cavity displays one of the lowest power thresholds for optical bistability among 2D PhC silicon devices for which a similar power-dependent analysis was carried out.\cite{Notomi2010, Zhang2013}

A high $Q$-factor and a small modal volume are not the only important requirements in view of applications. Many photonic structures of current interest, such as e.g. coupled cavities or coupled-resonator waveguides,\cite{Notomi2010} rely on spatial proximity between two cavities or one cavity and one waveguide, while in a longer-term perspective the density of optical elements will represent a key figure of merit of photonic circuits. For this, the spatial footprint of the PhC defect defining the cavity is a relevant figure of merit. The present cavity design is based on modified elementary cells of the PhC only up to five crystal periods away from the cavity center, namely half the value characterizing a typical ultrahigh-$Q$ design.\cite{Taguchi2011} The ability to produce compact-footprint cavities like the structure presented here thus also constitutes a major advance in view of an integrated photonic technology.

In summary, we demonstrated an ultrahigh-$Q$ $H0$ PhC nanocavity, fabricated from a silicon slab using an optimal design that we recently developed. The optimal cavity design was obtained by choosing to modify only a few technologically accessible variational parameters that preserve the small volume and small footprint of the cavity, and is characterizes by a theoretical quality factor $Q=1.7\times 10^6$. Our measurements result in an unloaded $Q$-factor of $450,000$. When accounting for the simulated mode volume, this corresponds to a $Q/V$-ratio exceeding $10^6(n/\lambda)^3$, ranking among the topmost values ever demonstrated in 2D PhC structures. The cavity displays optical bistability at a threshold power of $P_{cavity} = 13\mathrm{\mu W}$, (corresponding to a laser input power $P_{input} = 580 \mathrm{\mu W}$) i.e. one of the lowest reported for a silicon device. These features, combined with the compact design, make this cavity an ideal candidate element for silicon photonic integrated circuits. 

\begin{acknowledgments}
The authors acknowledge the financial support from the Swiss National Centre of Competence in Research Quantum Photonics and the Swiss National Science Foundation projects N\textsuperscript{\underline{o}} 200021\_134541 and 200020\_132407.  We also acknowledge Dario Gerace for fruitful discussions during the progress of the work.
\end{acknowledgments}



%

\end{document}